\documentclass[12pt,preprint]{aastex}
\usepackage{graphicx}
\usepackage{rotating}

\newcommand{\La}{Ly$\alpha$}
\newcommand{\Ot}{[O\,{\sc iii}]}
\newcommand{\Od}{[O\,{\sc ii}]}
\newcommand{\Ha}{H$\alpha$}

\begin{document}

\title{Search for z$\sim$6.96 \La\ emitters with Magellan/IMACS\thanks{This paper includes data gathered with the 6.5 meter Magellan Telescopes located at Las Campanas Observatory, Chile.} in the
  COSMOS field \thanks{Based on observations obtained at the Canada-France-Hawaii Telescope (CFHT), which is operated by the National Research Council (NRC) of Canada, the Institut National des Sciences de l'Univers of the Centre National de la Recherche Scientifique of France (CNRS), and the University of Hawaii. This work is based in part on observations obtained with MegaPrime/MegaCam, a joint project of CFHT and CEA/DAPNIA and in part on data products produced at TERAPIX and the Canadian Astronomy Data Centre as part of the Canada-France-Hawaii Telescope Legacy Survey, a collaborative project of NRC and CNRS.}}
\author{P.Hibon\altaffilmark{1}, S.Malhotra\altaffilmark{1}, J.Rhoads\altaffilmark{1}, C.Willott\altaffilmark{2}}
\altaffiltext{1}{School of Earth and Space Exploration,  Arizona  State University,  Tempe, AZ  85287}
\altaffiltext{2}{Herzberg Institute of Astrophysics, National Research Council, Canada}

\begin{abstract}
We report a search for z$\sim$6.96 \La\ emitters (LAEs) using a Narrow-Band filter,
centered at 9680\AA\, with the IMACS instrument on the
Magellan telescope at Las Campanas Observatory. %After defining our selection criteria, 
We obtain a
sample of 6 \La\ emitter candidates of luminosity
$\sim 10^{42}\mathrm{erg}\, \mathrm{s}^{-1}$ in a total area of 465 square arcmin corresponding to a comoving volume of $\sim 72000 Mpc^{3}$.\\
From this result, we derive a \La\ luminosity function (LF) at z$\sim$6.96
and compare our sample with the only z$\sim$6.96 \La\ emitter
spectroscopically confirmed to date \citep{Iye2006}. We find no evolution between the z=5.7 and z$\sim$7 \La\ luminosity functions, if a majority of our candidates are confirmed. %If one of the photometric candidates is spectroscopically confirmed as a z$\sim$7 LAE, it would confirm an evolution of the \La\ LF from z=6.5 to z$\sim$7.\\
Spectroscopic confirmation for this sample will enable more robust conclusions.
\end{abstract}

\section{Introduction}

Over the last decade, significant progress has been made in determining the processes of galaxy evolution using both ground- and space-based telescopes.  Currently, the limits of the observable universe are at z $\sim 6$, which corresponds to $\sim 90\%$ the age of the universe. At z$>$6, we are approaching the NIR domain : the sky is brighter and it is more challenging to detect the faint high redshift sources. Beyond this boundary lie the first ultraviolet (UV)-emitting sources, which ionized the majority of the hydrogen in the universe. Their detection will allow us to probe the era of reionization, after the ``Dark Ages''.  Galaxies formed at high redshifts play a key role in understanding how and when the reionization of the universe took place.  They also help constrain the physical mechanisms that drove the formation of the first stars and galaxies in the universe.  

%Over the last decade, significant progress in determining the processes
%of galaxy evolution has been made using the combined power of
%ground-based telescopes and the Hubble Space Telescope. The limits of
%the observable Universe have been pushed to a redshift of 6 which
%corresponds to looking back over about 90\% of the age of the
%Universe.
%Pushing the limits of the observable Universe beyond a redshift of 6
%by detecting the first ultraviolet-emitting sources, will allow us to
%probe the era of reionization,a few hundred million years after
%the end of the Dark Ages, when the first light-emitting objects, which
%ionized most of the content of the Univers, appear.\\
%The observations of high redshift galaxies play a key role for
%understanding the formation of the first stars and galaxies in the
%Universe, defining constraints on the physical mechanisms that drove formation
%and evolution of galaxies during the early universe, and for
%characterizing how and when the reionization of the Universe took place.\\

Starbursting galaxies can emit a large fraction of their ultraviolet luminosity in
the \La\ line. Because \La\ photons are resonantly
scattered in neutral hydrogen, even a small amount of dust can quench
this emission. Hence, selecting objects with strong \La\
emission lines is expected to reveal a set of objects in the early
phases of rapid star formation. These could either be young objects in
their first burst of star formation or evolved galaxies undergoing a
starburst due to a recent merger. Selecting galaxies with strong
emission lines also allows us to probe the high-redshift \La\ luminosity function (LF).\\
Once the Ly$\alpha$ LF is determined, it is then possible to infer the ionization fraction of the intergalactic medium (IGM) at different redshifts \citep{Malhotra2004, Stern2005, Furlanetto2006, Kashikawa2006, Ouchi2010}. The presence of neutral hydrogen in the IGM  can reduce the Ly$\alpha$ flux of galaxies, it is therefore clear that the Ly$\alpha$ LF is sensitive to the ionization fraction of the Universe. If we knew the intrinsic LF(z) of galaxies at each redshift, a deviation of the observed LF from this intrinsic distribution could be attributed to the attenuation by HI, and hence be used to infer the ionization fraction. In practice, the approach is to do proceed to a comparison of Ly$\alpha$ LF at different redshifts, since the LFs of Ly$\alpha$ emitters (LAEs) don't evolve much between z=3-5.7 \citep{Cassata2010, Malhotra2011}.
%\textbf{If Ly$\alpha$ emitters (LAEs) and Lyman Break Galaxies (LBGs) are not intrinsically different populations \citep{Shapley2003, Verhamme2008, Nagamine2010}, by looking at the evolution of their LFs, it will be possible to observe a change in the IGM neutral fraction.}
%If the Ly$\alpha$ LF is unchanged between two epochs, it therefore implies that the IGM neutral fraction changes little.
%\cite{LeDelliou2006} have performed simulations of the properties of LAEs using their semi-analytical model based on hierarchical galaxy formation. This model reproduces well the LFs of Ly$\alpha$ emitting galaxies from redshift 0 to 6.5 and allows to make predictions to higher redshifts, between 7 and 20. The main prediction of their model, concerning this proposal, is a likely moderate decline of the bright end of the LF of LAEs from z=6.5 to z$\sim$7 arising from the evolution of the mass distribution of dark matter halos. 
%If the LF may undergo limited evolution between z$=6.5 $ and z$=7$, the effects induced by the incomplete reionization of the IGM may play an important role in the evolution of the observed LF.\\

With ground-based telescopes, the detection of very distant objects requires observation of UV spectral signatures that have been redshifted into the visible spectrum.  The longer the wavelength of the observed Ly$\alpha$ line, the earlier the epoch at which we observe the galaxy, and the closer to the ``Dark Ages''.  Therefore, one way of searching for the most distant galaxies is to search for the redshifted Ly$\alpha$  emission at the longest possible wavelength.  However, this search is complicated by the presence of OH emission lines within the terrestrial atmosphere, at an altitude of $\approx 80km$.  This strong 
line emission limits the sensitivity of ground-based telescopes at near-infrared wavelengths. Fortunately, there are spectral intervals with lower OH-background that allow for a fainter detection limit from the ground.  %This is known as the narrow-band (NB) imaging technique.  

We use a custom-built filter, $NB9680$, centered at $\lambda=9680\AA\ $ and with a width of 90\AA to use one of the low-sky windows.
Narrow-band imaging is the most successful method to detect strong Ly$\alpha$ emission lines of galaxies, since it relies on a specific redshift interval as well as a selected low-sky background window.  Adapting the spectral width of this filter allows for maximum detection of light from the celestial objects at that spectral line, while minimizing the adverse influences of sky emission.  

%Over a thousand of LAEs have been photometrically selected and
%spectroscopically identified.\\

%Actual Statement Kashikawa, Iye, Hibon Tilvi, Castellano, Ouchi?\\
%To date, \cite{Kashikawa2006} has the biggest sample of z=6.5 \La\ emitters
%and \cite{Iye2006} has detected and spectroscopically confirmed the
%the highest \La\ emitter at z=6.96. Conclusions on LF from  \cite{Kashikawa2006}
%and \cite{Iye2006}. 
%%%%%%%%%%%%%%%

%Need to speak about Hu and Ota Ouchi.\\
%speak about 5.7, 6.5 : Hu, Kashikawa, Shimasaku, Ouchi, Westra, cassata
%samples
%evolution LF
%z=7
%Iye, Ota, Castellano
%results
%attempts at z=7.3 Ouchi, z=7.7, 8.8, and more : Cuby, Willis, Hibon, Tilvi ..
	The first  $z > 6$ LAEs was detected with the narrow-band (NB) technique at the 10m KeckII telescope \citep{Hu2002}.  This galaxy was spectroscopically confirmed to be at $z = 6.56$.  Over $1,000$ LAEs have been photometrically selected and spectroscopically identified in this way.  %Extensive observation have been done at $z = 5.7$ and $6.5$, two spectral domains free of sky lines in the optical spectrum.  
%PUT IN THE REST OF THE SURVEY STUFF HERE.  BE CAREFUL TO DEFINE LUMINOSITY FUNCTION (LF), VVDS, AND IGM BEFORE YOU USE THE ACRONYMNS.  ALSO, YOUR SPACING IS ALL OVER THE PLACE.  	
%The first z$>$6 found and spectroscopically confirmed was at z=6.56 by \cite{Hu2002} with the 10m Keck II telescope using NB technique.\\
%This observation method has yielded a consequent number of results. 
Extensive observations have been done at redshifts 5.7 and 6.5, two spectral domains free of sky lines in the optical spectrum, and different conclusions on the Luminosity Function are discussed by several groups \citep{Malhotra2004, Malhotra2006, Ouchi2008, Cassata2010, Ouchi2008,Ota2008, Iye2006, Kashikawa2006, Shimasaku2006}.
Several surveys have attempted to observe z$\sim$7.7 \citep{Hibon2009, Tilvi2010} and z$\sim$8.8 \citep{Cuby2007, Willis2008}, with no spectroscopic confirmation yet.\\
We present here a new NB imaging survey with the IMACS/Magellan telescope -- targeting $z = 6.96$ LAEs.  This paper first presents the data (Section 2.1) and the data reduction procedure (Section 2.2).  We then describe the method of selection and contamination of low-redshift interlopers for high redshift LAEs in Section 3.  We present the final sample of $z = 6.96$ LAEs and Ly$\alpha$ luminosity function at this redshift in Section 4. 

%We present here a new NB imaging survey with IMACS/Magellan telescope targeting z=6.96 LAEs.\\
%The outline of this paper is as follows. After  presenting the data
%and the data reduction procedure used in Section 2., we describe the
%method of selection for high redshift LAEs and the different
%low redshift interlopers who can contaminate this sample, in Section
%3. In Section4., we present the final sample of z=6.96 LAEs and the
%first attempt to build a \La\ Luminosity Function at z=6.96.\\
Throughout this study, we adopt the following cosmological parameters :
$H_{0}=70km.s^{-1}.Mpc^{-1}$, $\Omega_{m}=0.3$, $\Omega_{\Lambda}=0.7$ \citep{Spergel2007}.
 All magnitudes are AB magnitudes.

\section{Observations and Data Reduction}

\subsection{Observations}
The data were taken with the IMACS instrument (the Inamori-Magellan Area
Camera \& Spectrograph), installed at the
6.5m Magellan Baade telescope at Las Campanas Observatory. This
instrument offers two cameras: f/2 and f/4, with different imaging
scales and dispersions. We observed with the f/2 camera which delivers an image
of 27.4' diameter field at a scale of 0.2 arcsec per pixel.
To ensure background limited performance, each exposure lasted 15
min. After each exposure, the telescope was offset and a new exposure
was taken. The offset values, for each exposure, was randomly
chosen. \\
We targeted one COSMOS field (RA=10:00:29 Dec=02:12:21) covered by
 the Canada France Hawaii
Telescope Legacy Survey (CFHTLS) and the WIRCam Deep Survey (WIRDS :
PIs Willott \& Kneib)\footnote{Based on observations obtained with
  WIRCam, a joint project of CFHT,Taiwan, Korea, Canada, France, at
  the Canada-France-Hawaii Telescope (CFHT) which is operated by the
  National Research Council (NRC) of Canada, the Institute National
  des Sciences de l'Univers of the Centre National de la Recherche
  Scientifique of France, and the University of Hawaii. This work is
  based in part on data products produced at TERAPIX, the WIRDS
  (WIRcam Deep Survey) consortium, and the Canadian Astronomy Data
  Centre. This research was supported by a grant from the Agence
  Nationale de la Recherche ANR-07-BLAN-0228}. The total area of the
survey is 572 square arcminutes.
%EXPLAIN FILTER HERE!
%\subsubsection{2009}
We observed using the NB filter centered at 9680\AA\ ($NB9680$) on the 17th, 18th and 19th March 2009. 
During these observations, the conditions were good :  we obtained
19, 23, and 19 15-minute exposures during these nights, with a seeing
varying between 0.45'' and 0.8''.

%\subsubsection{2010}
We observed the field again on the 21st, 22nd and 23rd March 2010. Between the 2009 and 2010 observing season, the filter was replaced with a new one built to have identical bandpass specifications. %(The older filter was introducing some wavefront degradation, which became noticeable under the best seeing conditions. We shipped the filter back to the vendor, who furnished a new one with better transmitted wavefront.) We have noticed no 
There is no major difference between the two filters' realized bandpass, and we treat data from the two identically. 
%We observed with the same NB9680 filter than in 2009 during three nights in
%March 2010.
The night conditions were not as good as those in 2009 : the
seeing varied from 0.7'' to 1.3'' and we lost on average 2 hours per night due to instrument problems. We obtained 16, 19,
and 18 15-minute exposures in three nights.
The total exposure time for each epoch of data is given in Table \ref{table1}.

\subsection{NB Data Reduction}
\subsubsection{Data reduction}\label{dr}
To process this data, we use the package MSCRED/IRAF.
At the beginning of each night, we take 10 bias frames and 10 dome flats, we process the bias frames by overscan subtraction, trimming, and stacking to produce a master bias frame for each night.
%10 bias frames are taken at the beginning of each night. Once
%over-scanned and trimmed, the bias are combined per night.
%A similar process is apply to the 10 dome-flat frames taken for each
%night, before being normalized.
The science frames are bias subtracted and flat field corrected in the standard way. But the main difficulty was the fringing correction.\\
%EXPLAIN FRINGE, ORIGIN, EFFECT AND CORRECTION\\
%\textbf{p. 5, at top:  The discussion of illumination frame generation looks garbled to me, are you trying to describe a two-stage median smoothing algorithm here?  If so, do you really mean the first step used 8-12 pixel smoothing and the second used 13-20 pixel smoothing?  If so, what you really get is pretty similar to smoothing with a filter ~ 12*13 ~ 160 pixels on a side, which is then some 32 arcseconds.  The small smoothing scales quoted here would worry me as an illumination correction derived from the data with so little smoothing would end up dividing out real structure from the images--- the result would look funny.}

We see fringes produced by the interference of light reflected between
parallel surfaces in an instrument. %They appear in many detectors of
%visible and infrared light, among others in IMACS detector. 
%There are different fringe correction methods : by Fourier filtering,
%by wavelet analysis. 
All object frames do not always share the
same fringe pattern because flexure and variations in illumination
geometry can change its amplitude or period even on short
timescales. In order to correct each individual image for fringes, 
we produce a median image per night. An illumination frame is
also generated
for each science frame using a two-stage median smoothing algorithm, applied on each frame with a first stage filtering on a 16 pixel scale, the second on a 24 pixel scale. The net effect is comparable to a 384 pixel median filter corresponding to 76.8 arcsec.
%between 1.6'' -2.4'' for the initial one and 2.6''-4'' for the second one.

 We then produce a median of the illumination
frame for each night and correct the science frames for the illumination pattern. We subtract this illumination frame from the median of the science frame to
obtain a fringe pattern for each night. It is then essential to
find the optimal multiplicative scaling factor to use in removing the fringe pattern from each science frame, as we have long exposure frames and the
fringe pattern changes with time.
%sky subtraction\\
Once the frames are corrected for the fringes, we perform a sky
subtraction by subtracting a normalized median image of all science frames from each frame. %NEED TO BE CHECK.
%SKY BRIGHTNESS value per epoch\\
%mos\\
%stack for indivual nights and for the 3 nights\\
Finally, once the sky-subtracted frames are reconstructed as single extension images,
we stack them using the \textit{mscstack}/IRAF task.
We made stacks per night, per epoch and a combined epoch stack.\\

\textit{Astrometric calibration.}
We perform the astrometric calibration on the individual images before
the final stacking.
We set an initial WCS information in the header of the frames based on
the COSMOS catalog\footnote{This research has made use of the NASA/IPAC Infrared 
Science Archive, which is operated by the Jet Propulsion Laboratory, California 
Institute of Technology, under contract with the National Aeronautics and Space 
Administration.}. 
We need then to adjust the WCS parameters to
obtain a more precise alignment. For this purpose we use \textit{mscwcs} to
apply a first offset on the (RA,Dec) coordinates and we refine the
calibration by checking the alignment with \textit{msccmatch} task.%using msccmatch who check the alignment. 
%Depending on the rms of the positions, we then decide to accept the proposed
%correction by msccmatch.
We finally obtain an astrometry calibration for each individual images
with a precision of rms$\sim \pm0.1$ arcsec in both directions.\\

\textit{Photometric calibration.}
The photometric calibration of the CFHTLS data is based on the SDSS 
data for stars with 17$<$i'$<$21 and the
Megacam-SDSS color transformation equations of \cite{Regnault2009}.The precision
obtained in $u^\ast$, $g'$, $r'$, $i'$, $z'$ is between 0.03 and 0.02 mag.
As the $NB9680$ filter is included in the $z'$-band filter, we used the
MAG AUTO magnitude from the $z'$ band SExtractor catalog to calibrate our
$NB9680$ filter. We select 1048 non-saturated stars in the magnitude range 16$<z'<$20 to perform the calibration.
Considering the photometric error on the broad-band (BB) calibration, we obtain a
photometric calibration precise to 0.1 magnitude in $NB9680$.\\

\textit{Completeness.}
We estimate the limiting magnitude for the different bands by adding
200 artificial star-like objects per bin of 0.1 magnitude in blank
regions of the different stacked images. We then run SExtractor on the
image with the same parameters as previously used for object
detection. We repeat this procedure 40 times. 
The average count on 40 times of  the number of artificial stars retrieved in each
magnitude bin provides a direct measure of the completeness limit. We
report the 50\% completeness limit in Table~\ref{table1}.\\
We use this result to determine the Luminosity Function presented in Section~\ref{discussion}.

%\BEGINlt{figure}
%\centering
%\resizebox{\hsize}{!}{\includegraphics{NBc.pdf}}
%\caption{Completeness for the combined epochs NB9680 image.}
%\label{fig1}
%\end{figure}

\subsection{Broad Band Data}
\begin{figure}
\centering
\resizebox{\hsize}{!}{\includegraphics{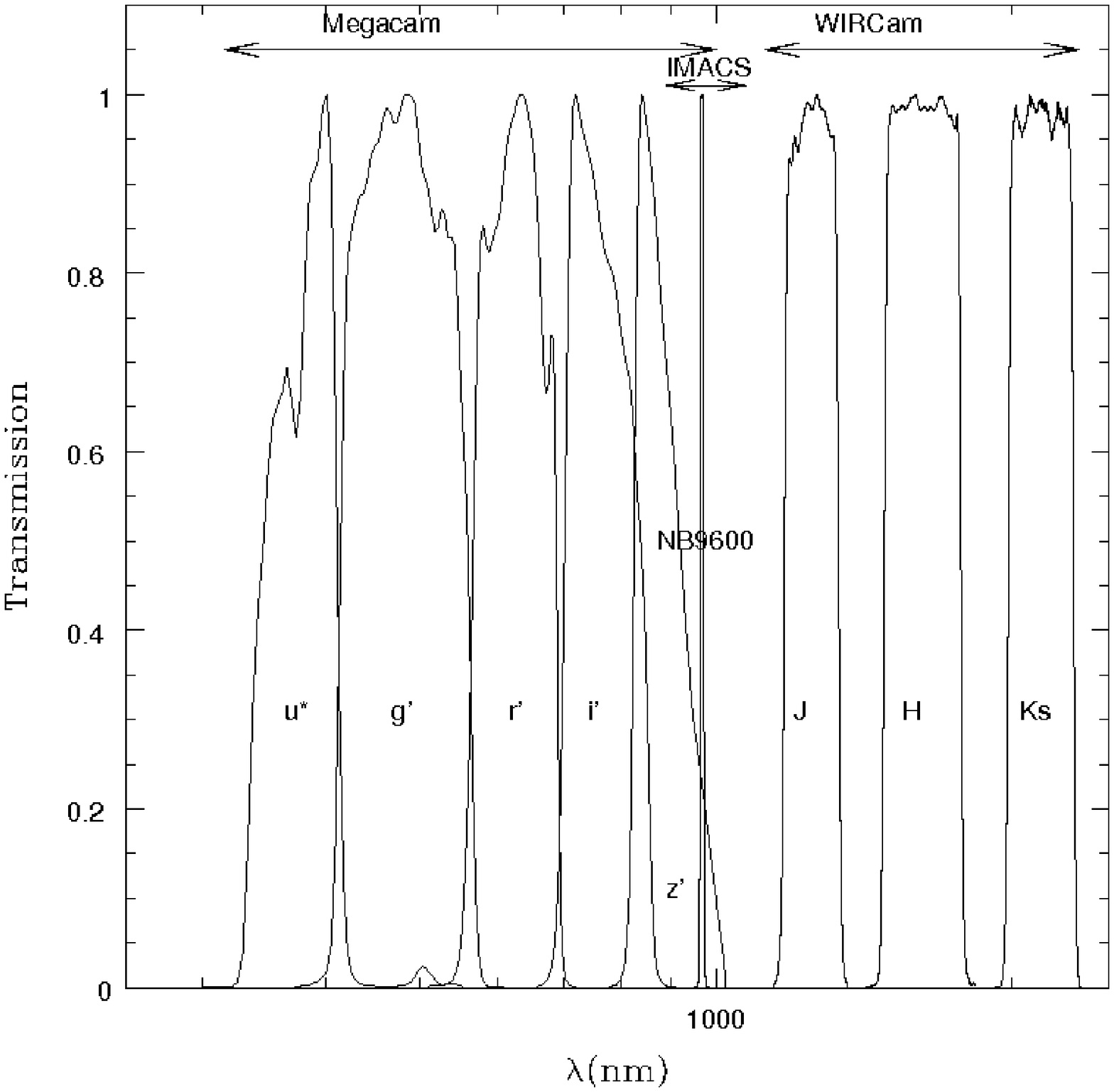}}
\caption{Transmission curves of the filters corresponding to the data
used in this paper. All transmissions include the response of the detector and are normalized to 100\% at maximum.}
\label{fig2}
\end{figure}

The CFHT-LS provides extremely deep optical imaging data for our
observed field. For the purpose of this study, we made use of the
T0006 release. The CFHT-LS data products are available from the CADC
archive to CFHT users and take form of image stacks in the $u^\ast$, $g'$, $r'$, $i'$, $z'$
filters and of ancillary data such as weight maps, catalogs etc.  The
spectral curves of the filter $u^\ast$, $g'$, $r'$, $i'$, $z'$ are similar to the ones of SDSS
filters.
In addition to the optical data, we also have the WIRDS survey, providing $J$, $H$, $Ks$.%in hand the Near
%Infrared deep Imaging data from the WIRCam Deep Survey (WIRDS).\\% This data products
%are available from CADC archive to CFHT users, as the optical data.\\
These optical data have been calibrated photometrically using the SDSS photometry
 and the NIR data using 2MASS photometry \citep{McCracken2010}.
Considering internal and external photometric error sources, the uncertainty 
on the optical and the NIR data photometry is $\sim$0.05 mag and 
$\sim$0.02 mag, respectively.\\
As the broad-band data and the narrow-band data do not have the same
pixel scale, we resample the broad-band data using the software
SWARP to obtain optical and NIR images with 0.2 arcsec/pixel.
The alignment in pixels is then verified using IRAF \textit{geomap/geotran}
tasks.
Our complete set of data is therefore scaled at 0.2 arcsec/pixel and
covers an area of 27.4' diameter. 
A summary of the observational data used in this paper is provided in
Table~\ref{table1}.
Figure~\ref{fig2} shows the transmission curves of the filters corresponding
to the multi-band data used in this study. 

\begin{table}
\caption{Observational data.} \label{table1}
\centering
\begin{tabular}{c c c c} \hline \hline
Instrument    & Band  & Integration            & Limiting \\ 
              &         &  time (hrs) & magnitude$^{\mathrm{a}}$\\ \hline 
MegaCam   &         $u^\ast$   &       12.5           &  26.6\\ 
MegaCam   &         $g'$       &       20.25          &  27.9\\
MegaCam   &         $r'$       &       37.6           &  27.6\\
MegaCam   &         $i'$       &       54             &  27.3\\
MegaCam   &         $z'$       &       40.1           &  26.4\\ \hline
IMACS        & $NB9680$ 1st epoch  &       15.25            &  25.\\
IMACS        & $NB9680$ 2nd epoch  &       13.25            &  25.$^{\mathrm{b}}$ \\
IMACS        & $NB9680$ combined   &       28.3             & 25.2$^{\mathrm{c}}$\\
WIRCam     &     $J$         &                        &  24.8\\ 
WIRCam     &     $H$         &                        &  24.9\\ 
WIRCam     & $K_{\mathrm s}$  &                         &  25\\ \hline
\end{tabular}
\begin{list}{}{}
\item[$^{\mathrm{a}}$] $5 \sigma$ magnitude limits in apertures 1\arcsec\
in diameter for MegaCam and WIRCam. These limits correspond to a 50\% 
completeness level.
\item[$^{\mathrm{b}}$] The filter used during 2010 observations has a better transmitted wavefront than the filter used in 2009. This explains the similar limiting magnitude between the two epochs although the observing conditions were different.
\item[$^{\mathrm{c}}$] By combining the two epochs of data, we should have expected a sensitivity increase of 0.37 mag. But the limit has increased by only 0.2 mag. This could be due to a systematic noise component like the fringes.\\
\end{list}
\footnote{The quoted limits [for the $NB9680$ filter] are based on the enclosed flux in a $1''$ diameter aperture.  The {\it total} fluxes and magnitudes for our candidates (e.g., fig.~\ref{fig3} and table~\ref{tab:cand}) include a substantial aperture correction (discussed in section~\ref{3.3}).}
\end{table}

%% In this section, we use  the \subsection command to set off
%% a subsection.  \footnote is used to insert a footnote to the text.


%% This section also includes several of the displayed math environments
%% mentioned in the Author Guide.

\section{Sample }
\subsection{Catalog generation and Selection}\label{cat}
We generate the catalogs using the software SExtractor \citep{Bertin1996}. We use the dual-image mode : the first image, for the detection, is settled as the combined $NB9680$ image, the second image, for measurement, corresponds to the resampled images from the optical and NIR bands.
%In order to be able to use this dual-image mode, we resampled the optical, single epoch NB and NIR images to the IMACS pixel scale using the software SWARP.\\
We choose to detect objects in 7 pixels above a threshold of 1.5$\sigma$, corresponding to a $NB9680 \sim 25.8$. The aperture used for the photometry is 1 arcsec diameter.

%\subsection{Selection}

%The useful solid angle of the survey-- the part used in candidate finding-- should be explained in the candidate selection part of the manuscript:  what did you throw out, why, and how many square arcminutes were ultimately used for finding candidates.
%The number I assume is the 465 or so that appears in the conclusions, but it should be explained earlier.
As the IMACS instrument is composed of eight chips, we see an increase of the noise in the inter-chip regions. 
%NEED TO BE CHECK!!!
%Using a similar formula than \cite{Duriscoe2007}, to estimate the sky brightness, we evaluate this parameter in an interchip region and in a more central part of the combined 2-epochs stack. We obtain a difference of \textbf{sky noise of a factor of 2.5 }between these two regions, the inter-chips one being the brightest one.
%We evaluate the sky noise in an interchip region and in a more central part of the combined 2-epochs stack. We obtain a difference of sky noise by a factor of 2.5 between these two regions, the inter-chips one being the noisiest one.
%need to evaluate the difference between in normal regions and in interchip regions.-> see tilvi method to get sy noise in mag.arcsec2.
The sky noise is higher by a factor 2.5 in the interchip regions.
We choose therefore to eliminate these regions from the area of the survey. From the total area covered by our survey, 572 square arcminutes, we obtain an effective area to search for z$\sim$6.96 LAE candidates of 465 square arcminutes.\\

Criterion\#1 : Since we have two epochs of data, we define
$NB9680$ selection criteria on individual epochs and on the combined images.
We select objects with a 3$\sigma$ detection in both of the individual
epoch images and a 5$\sigma$ detection in the combined $NB9680$ image. 59\% (=16150 objects) of the objects present in the initial catalog pass this criterion. This eliminates variable sources from the catalog.\\

Criterion\#2 : Due to the nearly complete absorption
of the flux shortward of \La\ by the intergalactic Hydrogen, we
should observe flux discontinuity at rest wavelength of 1216\AA, and observed wavelength of 9680\AA.% at redshifts greater than about 6. 
We are therefore searching for objects which are not detectable in
optical ($u^\ast$, $g'$, $r'$, $i'$) bands. A possible method is therefore to select objects with less than a
3$\sigma$ detection in filters blueward of the expected \La\ emission :
$u^\ast$, $g'$, $r'$, $i'$. \\
%The color break between the optical and $NB9680$
%filters is high and covers a wide spectral range. %However, for the
%CFHT-LS, the Terapix data center generated deep $\chi^2$ image
%combining the $g'$, $r'$ and $i'$ images. The precedent method can therefore be replaced as an unique selection criterion
%expressed as a less than 3$\sigma$ detection in the 
%$\chi^2$ image. 3\% of the previous sample responds to this criterion.\\
We use a $\chi^2$ image generated by combining $g'$, $r'$ and $i'$ CFHT data to obtain deep photometry of candidates in the combined optical bands. 3\% of remaining candidates are accepted on this basis.\\

\begin{figure}
\centering
\resizebox{\hsize}{!}{\includegraphics{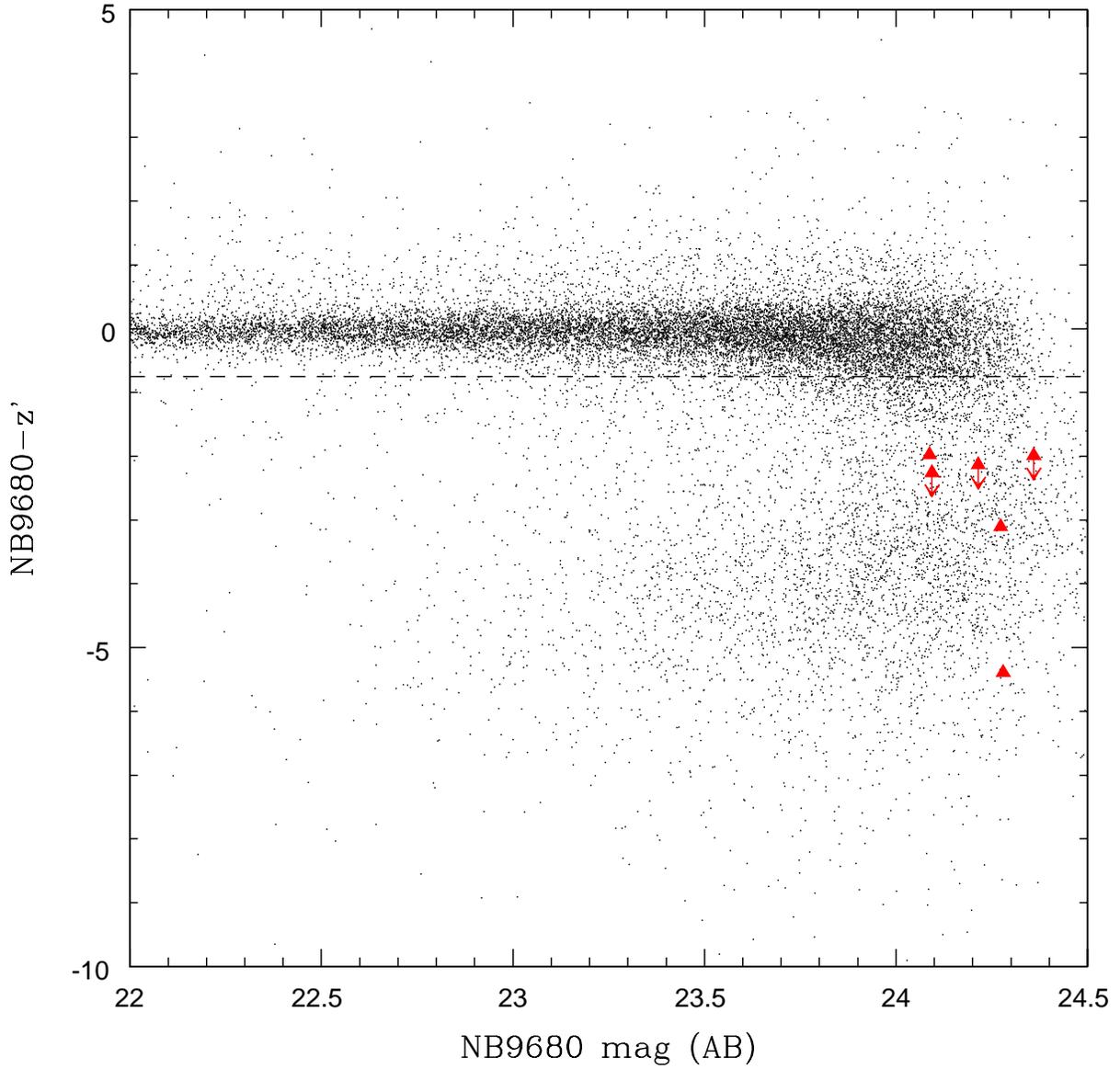}}
\caption{Color-Magnitude Diagram $z'$-$NB9680$ vs $NB9680$ showing the candidates
  obtained with our criterion $mag_{NB9680}-mag_{z'}<-0.75mag$ (plain red triangle). The other points with  $mag_{NB9680}-mag_{z'}<-0.75mag$ are ruled out by other criteria (blue flux, variability etc..).}% and the
  %criterion from \cite{Ota2008} $mag_{NB9680}-mag_{z'}<-1.72mag$ (plain blue circle). }
\label{fig3}
\end{figure}

Criterion\#3 : %The NB9680 filter used for this study is included in the broad z band
%filter. Although we are expecting an excess of flux in NB9680, it is yet
%possible to observe part of the continuum in the z band
%filter. Therefore, since we cannot reject objects present in this broad band
%filter, we select only objects with a 3$\sigma$ flux excess in NB9680
%comparing to $z'$ band. Moreover, 
Following the successful method of
\cite{Rhoads2001} used for the search of z=5.7 LAEs in the LALA field, we require
that more than 50\% of the $NB9680$ flux comes from an emission line for object selection, which
can be translated as $mag_{NB9680}-mag_{z'}<-0.75mag$.
%We also consider one of the $z'-NB9680$ color criteria used by \cite{Ota2008} \textbf{:$mag_{NB9680}-mag_{z'}<-1.72mag$ }.
This criterion is indicated in Figure \ref{fig3}, showing the
$z'-$$NB9680$ color versus $NB9680$ magnitude diagram. 88\% of the objects selected after the criterion\#2, pass this color restriction.\\

Criterion\#4 : To avoid selecting extremely red objects \citep{Cimatti2002}, we choose to set a different criterion : $mag_{NB9680}-mag_{J}<0$. 98\% of the objects selected after the criterion\#3 pass this color criterion.\\

Criterion\#5 : After applying carefully all the criteria presented above, we inspect
each candidate visually. % to confirm their point-like source aspect.
33\% of the objects inspected carefully are selected as serious candidates. the rest are rejected because they are near chip boundaries, or they are defects, etc.% ADD A  SENTENCE WHY HAPPEN TO THE OTHER ONES

The equations summarizing this selection are presented in the Table~\ref{tab:crit}.

%\textbf{We also realized a second selection by modifying our criteria 2 and 3 by the ones of \cite{Ota2008}.}%DESCRIPTION
%The equations summarizing these two selections are presented in Table~\ref{tab:crit}.
\begin{table*}
\caption{Table of the selection criteria.}
\label{tab:crit}
\centering
\begin{tabular}{c c c } \hline \hline
%\textbf{Name}   & $NB9680$ & Error & SNR ($NB9680$) & $z'$  & Error & $SNR (z')$  & $EW^{\mathrm{a}}$ (\AA)& $J$ & Error\\ \hline 
             & Our criteria \\ \hline%& \cite{Ota2008} criteria \\ \hline
Criterion\#1 & SNR($NB9680_{combined}$)$> 5\sigma$ \\%& SNR($NB9680_{combined}$)$> 5\sigma$  \\
             & SNR($NB9680_{2009}$)$> 3\sigma$    \\%& SNR($NB9680_{2009}$)$> 3\sigma$ \\
             & SNR($NB9680_{2010}$)$> 3\sigma$    \\%& SNR($NB9680_{2010}$)$> 3\sigma$ \\
Criterion\#2 & SNR($u^\ast$, $g'$, $r'$, $i'$)$< 3\sigma$ \\%& SNR($u^\ast$, $g'$, $r'$, $i'$)$< 3\sigma$ \\
%             &                                   & $mag_{i'}-mag_{z'}> 1.3 mag$\\   
Criterion\#3 & $mag_{NB9680}-mag_{z'}< -0.75mag$ \\%& $mag_{NB9680}-mag_{z'}< -1.72mag$   \\        
Criterion\#4 &  $mag_{NB9680}-mag_{J}< 0$ \\%& $mag_{NB9680}-mag_{J}< 0$  \\        
\end{tabular}
%\begin{list}{}{}
%\item[$^{\mathrm{a}}$] \textbf{In the restframe}
%\end{list}
\end{table*}

\subsection{Contaminants} \label{contaminants}
COSMOS photometric redshift catalog gives us a first insight on the low
redshift emitters contaminating our high redshift candidates sample.\\

\textit{Transient objects.}
Our strategy of observing the same field during two different years, and our requirement that the eligible candidates have to be detected in each epoch data  within 3$\sigma$ level, allow us to rule out 
the contamination of our sample by transient objects  such as supernovae, which would have appear in only one epoch of data. This strategy likewise prevents contamination by slow-moving solar system objects.\\
%Our observation strategy, consisting in observing the same
%field during two different years, and our selection criterion, defined
%as the fact that the eligible candidate have to be detected in each
%epoch data within \textbf{3}$\sigma$ level, allow us to rule out 
%the contamination of our candidates sample by
%transient objects such as supernovae, which would have appear in only
%\textbf{one epoch} of data. Moreover, to stack data acquired over long time
%spans, removes automatically slowly moving solar-system objects.\\
%\textbf{2- Supernovae}

\textit{ L-T dwarfs stars.}
Following the method described in \cite{Hibon2009}, we determine the
expected number of L and T-dwarfs present in this survey. From the
Figure 9 of \cite{Tinney2003}, representing the relation between the
absolute magnitude and the spectral types of late-type dwarf galaxies,
we find that we could detect L-dwarfs up to a distance of 871 to 3630pc
and T-dwarfs up to a distance of 400 to 1260pc, depending on spectral
type, from the coolest to the warmest. \\ 
%(DONE WITH Mlim-Z=26:NEED TO BE CHECK).\\
This field is located at high galactic latitude. Our sensitivity to L-
and T-dwarfs is then extended beyond the scale height of the
Galactic disk. However, the scale height applicable to L-, T-dwarfs is
truncated at 350pc \citep{Ryan2005}. We estimate therefore a sample
volume of $\sim 692pc^{3}$. Considering a volume density of L- and
T-dwarfs of a few $10^{-3}pc^{-3}$, we expect less than one L-,
T-dwarf in our field.\\
In addition, L-,T-dwarfs have $NB9680$-J$>$0  so they would have fail Criterion\#4 of our selection.\\ %we can have NB9680-J$<$0 to be certain of our selection.\\

Lower redshift galaxies with strong emission lines should be actively star forming galaxies with blue continuum emission. Looking for such a continuum allows us to identify these lower-redshift line emitters, unless their equivalent qidth is very large.\\
\textit{Foreground emitters.}
We estimate the minimum observed equivalent width a foreground line emitter would require to be selected with our criteria using the formula from \cite{Rhoads2001}:
\begin{equation}
EW_{min} \sim \left(\frac{f_{NB}}{f_{BB}}\right) \Delta\lambda_{NB} = \left[\frac{5\sigma_{NB}}{3\sigma{BB}}-1\right] \Delta\lambda_{NB}
\end{equation}
with $f_{NB}$ and $f_{BB}$ the flux in $NB9680$ and $g'$ band respectively, $\Delta\lambda_{NB}$ the width of the $NB9680$ filter, $\sigma_{NB}$ and $\sigma_{BB}$, the flux uncertainties in $NB9680$ and $g'$ band respectively.
We obtain therefore an $EW_{min}\sim 1545\AA$ in observer frame. Foreground line emitters would then require an observed equivalent width $EW_{min}\ge 1545\AA$ to contaminate our \La\ selection sample.\\
For guidance, this observed equivalent width corresponds to a rest-frame equivalent width of $EW_{rest} \ge 1051\AA$ for \Ha\ emitters at z$\sim$0.47, of $EW_{rest} \ge 792\AA$ for \Ot\ emitters at z$\sim$0.95 and of $EW_{rest} \ge 594\AA$ for \Od\ emitters at z$\sim$1.6. In the following studies, we used the observed equivalent width to estimate the number of emitters present in the survey and possibly contaminating our high redshift sample.

1- \Ha\ at z$\sim$0.47\\
We first estimate the fraction of \Ha\ emitters with $EW_{obs} \ge 1545\AA$ from Figure 2 of \cite{Straughn2009}. Fewer than 1.2\% of \Ha\ emitters at z$\sim$0.27 from \cite{Straughn2009} sample would have such an equivalent width.
We then evaluate the number of \Ha\ emitters at z$\sim$0.47 present in our
survey using the luminosity function from Figure 14 of \cite{Tresse2003}. 
We find that $\sim$ 12 \Ha\ emitters can be present in our $NB9680$ combined image. An upper limit of the number of \Ha\ emitters at z$\sim$0.47 passing our criteria is then 0.15. Considering the \Ha\ luminosity function of \cite{Geach2010}, we find that $\sim$ 52 \Ha\ emitters can be present in our survey. An upper limit of the number of \Ha\ emitters based on \cite{Geach2010} is therefore 0.62.
Thus \Ha\ emitters are not serious contaminants.
 %These foreground line emitters can therefore not contaminate our candidates selection.
% Using Figure 6 
%from \cite{Ilbert2005}, we estimate that we expect a R-band
%magnitude of $\sim$ 23.6 and an I-band magnitude of $\sim$ 23.3 for
%these emitters. These magnitudes expected for \Ha\ emitters are
%approximately 4 magnitudes brighter than the detection limit of our
%R-band and I-band images.\\

2- \Ot\ at z$\sim$0.95\\
We apply the same method for the \Ha\ emitters to estimate the number of \Ot\ emitters in our survey and contaminating our selection. 
From Figure 2 of \cite{Straughn2009}, we estimate the fraction of \Ot\ emitters at $EW_{min}$ to be one out of 136. We obtain therefore an upper limit of 0.74\%. 
\cite{Kakazu2007}\Ot\ emitter sample is covering a wide range of rest-frame equivalent width up to $EW_{rest}\sim 1000\AA$. As the rest-frame EW of the \Ot\ emitters that could contaminate our high redshift sample is around 792.1\AA\, we are therefore able to use their sample to estimate the possible number of \Ot\ emitters passing through our selection criteria.
From the luminosity function Figure 13 of \cite{Kakazu2007}, 48 \Ot\ emitters could be present in our survey. Applying the upper limit to this number, we find that a maximum of 0.35 \Ot\ emitters at z$\sim$0.95 could have high enough $EW$ to be selected. %have been selected.
%Using Figure 6 from \cite{Ilbert2005}, we estimate that we expect a $r'$ band
%magnitude of $\sim$ 25.7 and an $i'$ band magnitude of $\sim$ 25.2 for
%these emitters. These magnitudes expected for \Ot\ emitters are
%more than 1 magnitude brighter than the detection limit of our
%$r'$ band and $i'$ band images. The selection criterion applied to the optical bands rules out these objects for contaminating our candidates sample.\\
%We estimate the minimum equivalent width necessary to pass our
%selection criteria. Using the formula presented in \cite{Rhoads2001},
%we found that foreground emitters need an equivalent width at least
%equal to ...\AA\. Considering this value for the \Ot\ emitters, and the
%Figure 2 of \cite{Straughn2009}, we found that .... \Ot\ emitters could
%be present in the survey. \\

3- \Od\ at z$\sim$1.6\\
We apply the same method to estimate the number of \Od\ emitters in our survey and contaminating our selection.
From \cite{Straughn2009}, we obtain an upper limit of 3.3\%.
Using the luminosity function presented in Figure 5 of \cite{Rigopoulou2005}, 45 \Od\ emitters at  z$\sim$1.6 can be detected in our survey. However a maximum of 1.5 of these objects can pass through our selection criteria.
%Using Figure 6 from \cite{Ilbert2005}, we estimate that we expect a $r'$ band
%magnitude of $\sim$ 22.6 and an $i'$ band magnitude of $\sim$ 22.1 for
%these emitters. These magnitudes expected for \Ot\ emitters are
%approximately 4 magnitudes brighter than the detection limit of our
%$r'$ band and $i'$ band images. The selection criterion applied to the optical bands rules out these objects for contaminating our candidates sample.\\
%We use the equivalent width distribution in function of redshift shown
%in Figure 1 of \cite{Teplitz2003}, to evaluate the rest-frame
%equivalent width of the \Od\ emitters at z$\sim$1.5-1.6 at
%$EW_{rf}($\Od$)\sim30-40$\AA\. We then report this value in the Figure 2
%of \cite{Straughn2009} showing the number of \Od\ emitters at z$\sim$1 in function of the rest-frame EW. Assuming no dramatic evolution of the rest-frame
%EW between z$\sim$1 and z$\sim$1.6, 8 \Od\ emitters are expecting in
%their volume, which, considering our volume, corresponds to 0.8 \Od\
%emitters. Our candidates sample is therefore not contaminated by \Od\
%emitters at z$\sim$1.6.\\

\subsubsection{False Detections}\label{fd}
In order to estimate the number of false detections that could pass our selection criteria, we create an inverse $NB9680$ image by multiplying the combined $NB9680$ image by -1. We then apply the same selection method and criteria and we did not find any candidates.

\subsubsection{Comparison with COSMOS redshifts}
We verify our sample of candidates by cross-correlating
this catalog with the photometric redshifts catalog from the COSMOS\footnote{This research has made use of the NASA/IPAC Infrared 
Science Archive, which is operated by the Jet Propulsion Laboratory, California 
Institute of Technology, under contract with the National Aeronautics and Space 
Administration.} field. This catalog covers a redshift 
range up to z$\sim$5.2 and a magnitude range from $z'\sim$18 to $z'\sim$25. None of the catalog's objects matches our candidate sample.
Since most of the case against foreground emitters is discussed in the previous sections and a consistency verification performed, we can therefore conclude that it is very unlikely that our z$\sim$6.96 
LAE candidate sample is contaminated by low-redshift interlopers.\\% or artefacts.\\
A more detailed study of the foreground emitters (\Ha\ at $\sim$0.47,
\Ot\ at $\sim$ 0.95 and \Od\ at $\sim$ 1.6) will be presented in a
forthcoming paper.\\

\subsection{Final sample}\label{3.3}
\begin{figure}
\centering
\resizebox{\hsize}{!}{\includegraphics{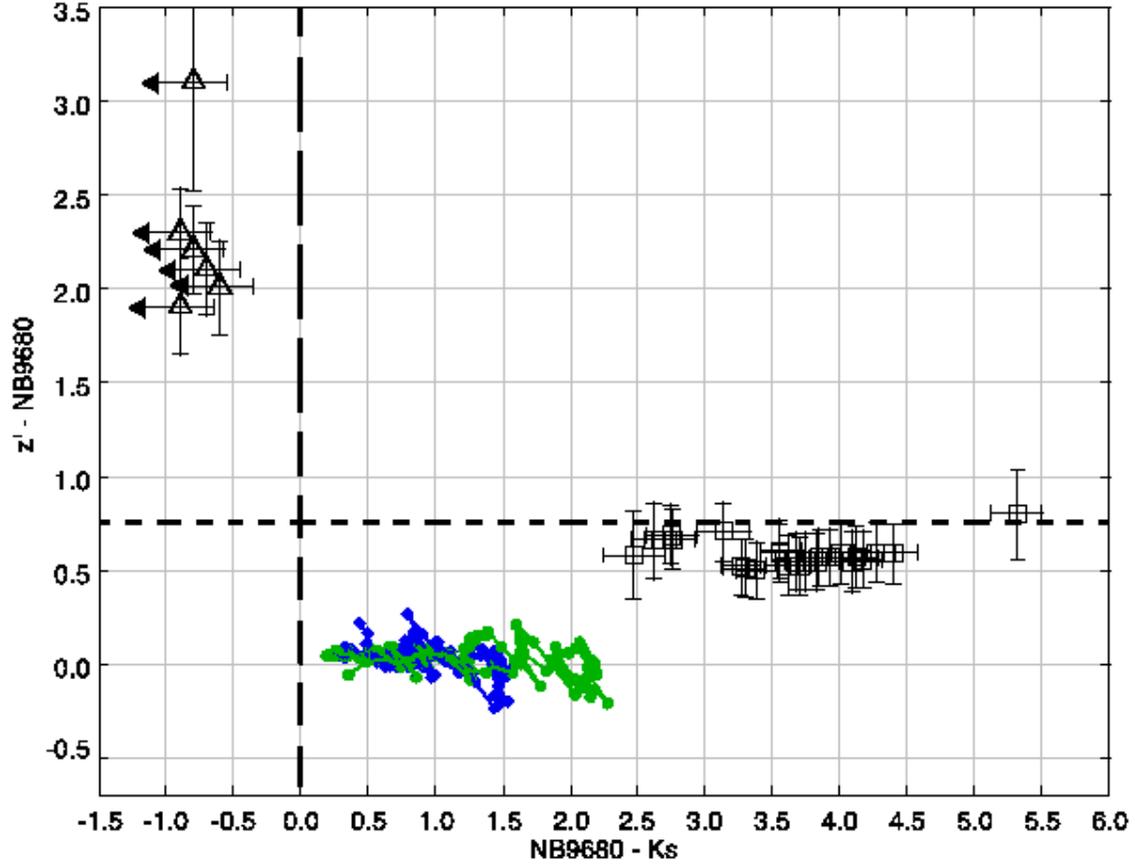}}
\caption{Color-Color Diagram $z'$-$NB9680$ vs $NB9680$-Ks showing our criteria $mag_{NB9680}-mag_{z'}<-0.75mag$ (short dashed line) and $mag_{NB9680}-mag_{Ks}< 0mag$ (long dashed line). The empty triangle symbol correspond to the LAE candidates. The empty squares represent the L-T-Dwarfs from \cite{Dahn2002}. The green track represents early-type galaxies and the blue track dusty galaxies generated with \cite{Bruzual2003}.}
\label{fig4}
\end{figure}

%how many\\
%certainty on quality and number\\
%characteristics : EW from WIRCam data formula etc...\\
%table with magnitude, SNR, EW, coordinates\\
%thumbnails\\
Our final sample contains 6 z$\sim$6.96 LAE candidates over the range of $NB9680$=24.1-24.4 and SNR ($NB9680$)=5.6-7.3, as described in Table~\ref{tab:cand}.
In order to derive a luminosity function independent of the photometry aperture, we compare the automatic aperture magnitude, the 1'' aperture magnitude and the isophotal magnitude for unsaturated objects. We then correct the 1'' aperture magnitude from the difference found between the different magnitude types. The aperture corrected magnitudes are presented in Table~\ref{tab:cand}.\\
Because our photometric calibration was based on matching SExtractor's automatic aperture magnitudes to the $z'$ filter photometric catalog from the COSMOS porject, this procedure should provide an unbiased estimate of the total (aperture-corrected) AB narrowband magnitudes for our objects. This approach also provides some robustness to crowding, thanks to the relatively small (1'') apertures used for color measurements described in Section~\ref{cat}.\\
We show our final sample in the $NB9680 - Ks$ vs $z' - NB9680$ diagram represented in Figure~\ref{fig4} with the different possible contaminants described above. The colors for L- and T- Dwarfs have been computed using the L- and T- Dwarfs library from \cite{Dahn2002}. We used GALAXEV \cite{Bruzual2003} to model the color tracks of early and dusty galaxies using the Padova 1994 evolutionary tracks with a Salpeter IMF.

\begin{figure*}
\centering
\includegraphics[width=16.cm]{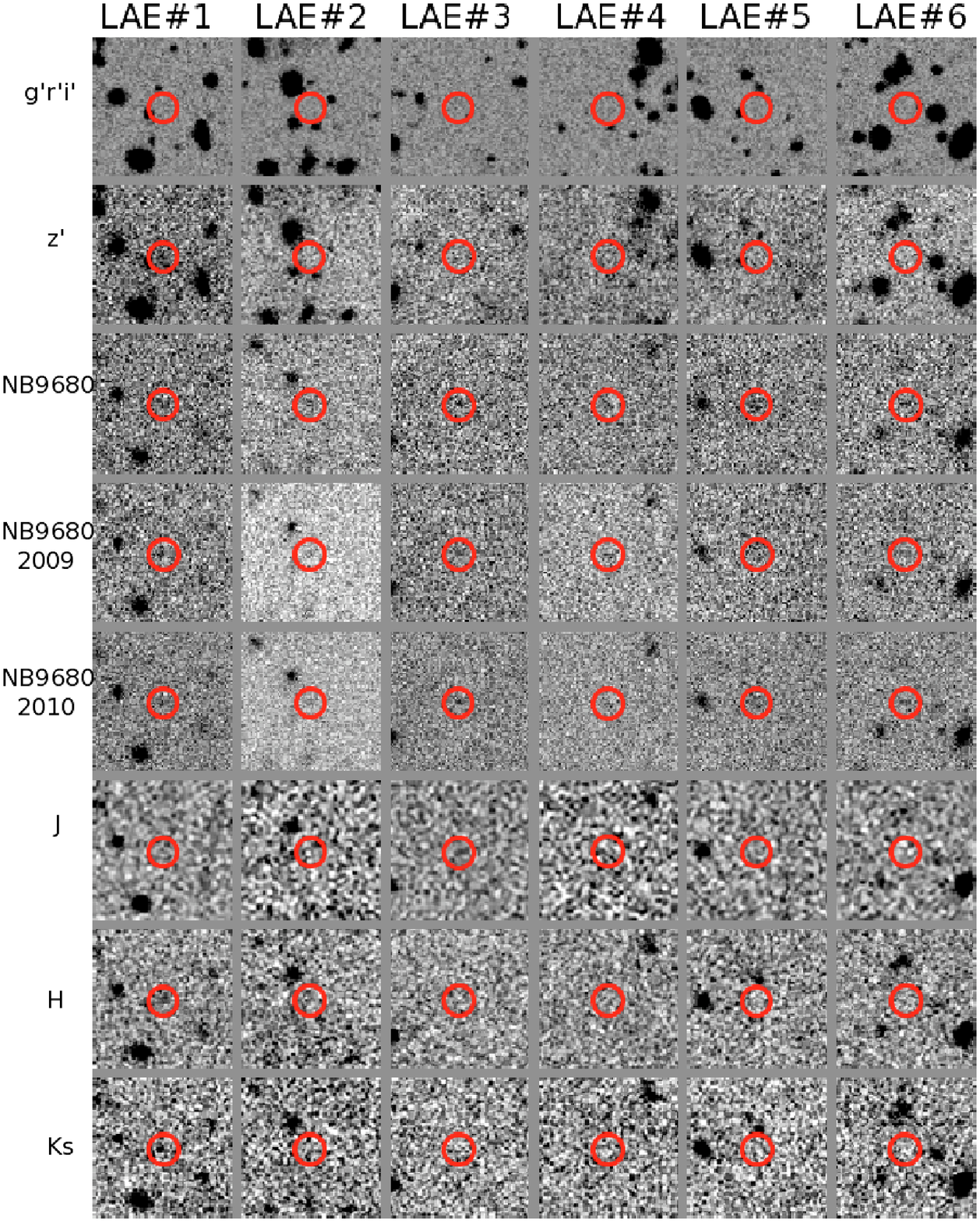}
\caption{Thumbnail images of all candidates listed
in Table~\ref{tab:cand}. Each window is $15''\times 15''$.  Objects names and passbands
are located above and to the left of the thumbnails, respectively.}
\label{fig:thumbnails}
\end{figure*}

We also report in Table~\ref{tab:cand} the lower limits of the rest-frame
equivalent widths ($EW$) derived from the photometric data following \cite{Malhotra2002}, defined as:
%SENTENCE ABOUT BRIGHT IN Ks.. MAGNESIUM POSSIBLE AGN...
 %This object could be therefore a high redshift object with a possible MgII line present in Ks band.
\begin{eqnarray}
EW_\mathrm{rest}(\AA) = \left(\frac{f_{NB9680} \Delta\lambda_{Z}- f_{Z} \Delta\lambda_{NB9680}} {f_{Z} - f_{NB9680}} \right) \times \frac{1}{1 + z}
\label{eq:ew}
\end{eqnarray}

where $f_{NB9680}$ is the observed flux in the narrow-band combined image, $f_{Z}$ is the observed flux in the $z'$ broad-band image, 
$\Delta\lambda_{NB9680}$  and $\Delta\lambda_{Z}$ are the width of the $NB9680$ filter (90\AA) and the $z'$ band filter (928\AA) respectively. \\  
%Considering the width of the NB9680 filter, we assume $f_{NB9680}\sim f_{Ly\alpha}$. As the NB9680 filter is covered by the $z'$ band filter, we also verified that the flux in $z'$ band is not "contaminated" by the \La\ line flux. \\
Our first object, LAE\#1 in Table~\ref{tab:cand}, present a detection in Ks band (see Figure~\ref{fig:thumbnails}). This could be due to a very red continuum slope. However, looking at the Spitzer/IRAC data in 3.5$\mu$m and 4.8$\mu$m, none of the LAE candidates are detected. Alternatively, it could be due to the presence of another line, such as MgII.
Looking at the UKIRT/WFCAM2\footnotemark[\value{footnote}] data in J band ($J_{50\%}=25.6$, AB, 5$\sigma$) for all our high redshift candidates, LAE\#1 is detected in this data with a magnitude of 24.8 (AB) and a SNR(J)$\sim$ 10. From the HST/NICMOS\footnotemark[\value{footnote}] data available in H band ($H_{50\%}=26.7$, AB, 5$\sigma$), LAE\#3 is detected with a magnitude of 26.5 and a SNR(H)$\sim$4.\\
These detections in different broad bands confirm the reliability on these candidates. Although the other candidates have a strong single band detection, the tests realized in Paragraph ~\ref{fd} confirm that they are emission line objects. The remaining
candidates are based on a $\ge 5\sigma$ significant single-band detection in the narrow band filter.
For Gaussian statistics, the false positive probability at $\ge 5\sigma$ is $3\times 10^{-7}$, while
our survey area contains $\sim 0.5 \times 10^7$ independent resolution elements (based on
$0.65''$ seeing).  The number of noise spikes entering the sample should thus be $\sim 1.5$, comparable
to the expected number of foreground emitters.  Non-Gaussian noise could increase this number, but the
absence of detections in the negative-image test (see Section~\ref{fd}) support the conclusion that $5\sigma$
noise spikes are not a major contaminant of our sample.

Three out of six of our LAE
photometric candidates are not detected in the $z'$ band and we
therefore use the detection limit in this band, deriving in turn lower
$EW$ limits. 

%\begin{table*}
\begin{sidewaystable}
\caption{Table of the z$\sim6.96$ LAE candidates.}
\label{tab:cand}
\centering
\begin{tabular}{c c c c c c c c c c c c} \hline \hline
\small{Name}   & \small{Ra} & \small{Dec} & \small{$NB9680$} & \small{Error} & \small{SNR ($NB9680$)} & \small{$z'$}  & \small{Error} & \small{$SNR (z')$}  & \small{$EW^{\mathrm{a}}$ (\AA)}& \small{$J$} & \small{Error}\\ \hline 
\small{LAE\#1}  & \small{10:00:46.846} & \small{02:10:16.01} & \small{24.1} & \small{0.2}   & \small{7.3}  & \small{26.0}     & \small{0.16} & \small{6.6} &\small{$\sim$41}  & \small{$24.8^{\mathrm{b}}$} & \small{0.15}\\        %8241  %8
\small{LAE\#2}  & \small{10:00:39.104} & \small{02:03:02.55} &  \small{24.1} & \small{0.18}  & \small{7.2}  & \small{$>$26.4}  &\small{ ...}  &\small{ ... }&\small{$>$61}     &\small{ $>$24.8} & \small{...}\\        %13833 %13833
\small{LAE\#3}  & \small{09:59:50.991} & \small{02:12:19.13} &  \small{24.2} & \small{0.2}   & \small{6.4}  &\small{$>$26.4}&\small{ ... }& \small{2}   &\small{$>$47} & \small{$>$24.8} &\small{ ...} \\        %7067  %8241
\small{LAE\#4}  &\small{10:00:58.529}& \small{02:12:56.51} &  \small{24.2} & \small{0.18}  & \small{6.1}  & \small{$>$26.4}  &\small{ ...}  & \small{...} &\small{$>$47}     & \small{$>$24.8} &\small{ ...} \\        %1     %1
\small{LAE\#5}  & \small{10:00:42.528} & \small{02:11:30.39}&  \small{24.3} & \small{0.19}  & \small{6.1}  & \small{$>$26.4}&\small{... }& \small{0.2} &\small{$>$48} & \small{$>$24.8} &\small{ ... } \\        %8     %7067 
\small{LAE\#6}  & \small{10:00:37.940} & \small{02:11:59.28} &  \small{24.4} & \small{0.2}   & \small{5.6}  & \small{$>$26.4}  &\small{ ...}  & \small{...} &\small{$>$41}     & \small{$>$24.8} &\small{ ... }\\ \hline %8767  %8767
%aper magnitude
\end{tabular}
\begin{list}{}{}
\item[$^{\mathrm{a}}$] In the restframe
\item[$^{\mathrm{b}}$] from UKIRT/WFCAM2 data
\end{list}
\end{sidewaystable}
%\end{table*}

\section{Discussion} \label{discussion}
\textit{Variance.}
 Two sources of variance can be involved in a such high redshift study
: the Poisson variance and the fluctuations in the large scale
distribution of the
galaxies, also called the cosmic variance. 
We used the on-line calculator\footnote{http://casa.colorado.edu/$\sim$trenti/CosmicVariance.html} by \cite{Trenti2008} to estimate the cosmic variance. This calculator requires numerous parameters such as the area of the survey, the mean redshift, the reshift interval, but also the completeness value and several cosmological parameters.
We obtained, for our sample, a value of 54\% for the cosmic variance.
%The on-line
%calculator\footnote{http://casa.colorado.edu/$\sim$trenti/CosmicVariance.html}
% by \cite{Trenti2008} gives a value
%of this cosmic variance, 54\% for our sample.\\
The Poisson noise is estimated to 58\%. The 54\% uncertainty from the cosmic variance and the 58\% from Poisson statistics are comparable and we therefore consider both in our error estimation.
%Considering the limited number of objects in our sample and the large 
%comoving volume of our survey, our results are more limited by Poisson 
%noise -- $\sim$58\% for the 6 objects -- than by clustering.

\begin{figure}
\centering
%\resizebox{\hsize}{!}{\includegraphics{lf2_AC_mod.pdf}}
\resizebox{\hsize}{!}{\includegraphics{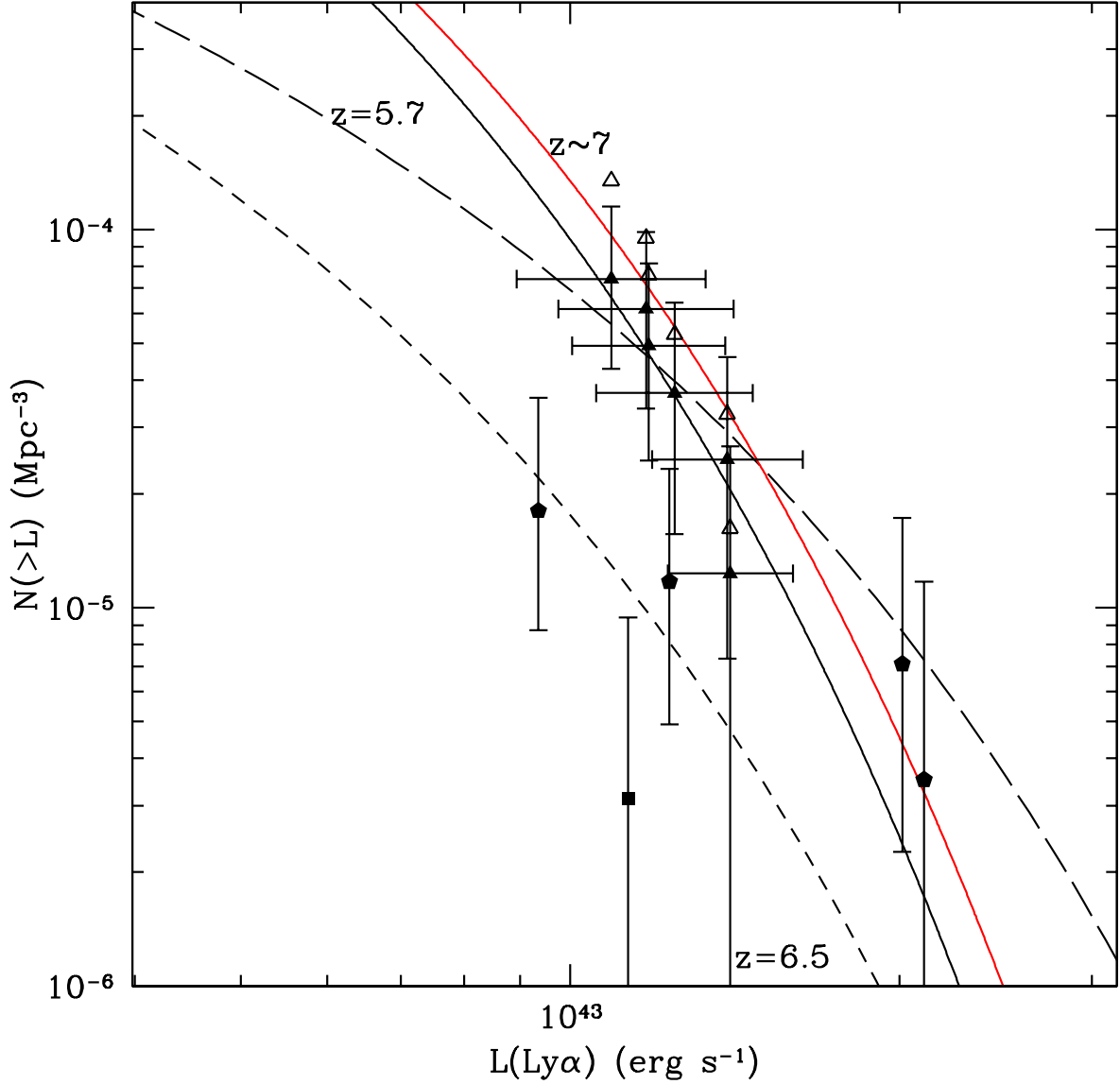}}
\caption{Best-fit Schechter function for the cumulative z$\sim$7 \La\ luminosity function. The red solid line is the best-fit z$\sim$7 LF with completeness correction. Our candidates are represented as triangles (plain: not corrected for completeness, empty : with completeness correction). The error bars represent the Poisson errors and are identical for plain and empty points. The z$\sim$6.96 LAE from \cite{Iye2006} is the square, \cite{Ota2010} candidates are pentagons. Also represented here is the z=6.5 \La\ LF from \cite{Ouchi2010} and the z=5.7 \La\ LF from \cite{Ouchi2008}.}
\label{lfus}
\end{figure}

\textit{Luminosity Function.}
%Following \cite{Taniguchi2005}, we assume that $\sim$70\% of the
%narrow-band flux comes from the \La\ line. We therefore apply a
%correction factor during the conversion from NB9680 magnitudes to \La\
%fluxes. \\
 Our $NB9680$ filter being quite narrow (FWHM$\sim$90\AA), we assume that the
narrow-band flux is entirely coming from the \La\ line. %Moreover, the data are not corrected for completeness. \\
We fit to the \La\ luminosity function
of this z$\sim$6.96 LAE sample, a Schechter function, $\Phi(L)$, given by
\begin{equation}
\Phi(L)\mathrm{d}L=\Phi^{*}\left(\frac{L}{L^{*}}\right)^{\alpha}\mathrm{exp}\left(-\frac{L}{L^{*}}\right)\frac{\mathrm{d}L}{L^{*}}
\end{equation}
in order to compare with previous high redshift works
\citep{Ouchi2010, Hibon2009, Tilvi2010, Ouchi2008, Ota2008, Kashikawa2006, Malhotra2004}.
Considering the small number of candidates in our sample, we choose to
fit two out of three of the Schechter function parameters. %Following
%\cite{Hibon2009, Ouchi2008},
We set the faint end slope of the
luminosity, $\alpha$, to $\alpha=-1.5$, and determine $\Phi^{*}$ and
$L^{*}$ by $\chi^{2}$ minimization. We decide to obtain a best-fit Schechter functions for the z$\sim$7 cumulative luminosity function, which has been derived by considering only our photometric candidates.\\
%We use the completeness result of Section\ref{dr} to determine the \La\ Luminosity Function represented in red in Figure\ref{lfus}.
%\textbf{In Figure~\ref{lfus}, we present the \La\ Luminosity Function not corrected for the detection incompleteness as a black solide line. We then correct the luminosity of our objects from the completeness result of Section~\ref{dr}, and we realize a new Schechter fit for the z$\sim$7 \La\ Luminosity Function corrected from incompleteness, presented as a red solid line in Figure~\ref{lfus}.
In Figure~\ref{lfus}, we present the \La\ Luminosity Function not corrected for detection incompleteness as a black solid line. We then use the completeness result of Section~\ref{dr} to correct the luminosity of our objects, and find a new Schechter fit, presented as a red solid line in Figure~\ref{lfus}.  This is the z~7 \La\ Luminosity Function corrected for incompleteness.

\begin{table}
\caption{Best fit Schechter LF parameters for $\alpha = -1.5$}
\label{tab:fitlf}
\centering
\begin{tabular}{c c c} \hline \hline
Redshift & $\mathrm{log}(L^{*} (\mathrm{erg}\, \mathrm{s}^{-1}))$ &
$\mathrm{log}(\Phi^{*}(\mathrm{Mpc}^{-3}))$ \\ \hline
%6.96$^{\mathrm{(1)}}$ &   $42.20^{+0.1}_{-0.2}$ & $-2.01^{+0.15}_{-0.2}$  \\
%6.96$^{\mathrm{(1)}\star}$ & $42.26^{+0.1}_{-0.15}$ & $-2.01^{+0.15}_{-0.15}$\\
%6.96$^{\mathrm{(1)}}$ &   $42.56^{+0.1}_{-0.2}$ & $-2.01^{+0.15}_{-0.2}$  \\
%6.96$^{\mathrm{(1)}\star}$ & $42.57^{+0.1}_{-0.15}$ & $-2.02^{+0.15}_{-0.15}$\\
%6.96$^{\mathrm{(2)}}$ &   $42.40^{+0.1}_{-0.2}$ & $-2.80^{+0.1}_{-0.1}$  \\
%6.96$^{\mathrm{(2)}}$ &   $42.39^{+0.1}_{-0.2}$ & $-2.01^{+0.1}_{-0.1}$  \\
%6.96$^{\mathrm{(2)}\star}$ & $42.39^{+0.1}_{-0.1}$ & $-2.01^{+0.15}_{-0.1}$ \\
6.96$^{\mathrm{(1)}}$ &   $42.56^{+0.1}_{-0.2}$ & $-2.01^{+0.15}_{-0.2}$  \\
%6.96$^{\mathrm{(1)}\star}$ & $42.6^{+0.1}_{-0.15}$ & $-2.01^{+0.15}_{-0.15}$\\
6.96$^{\mathrm{(2)}}$ & $42.8^{+0.12}_{-0.14}$  & $-3.44^{+0.20}_{-0.16}$ \\
6.5$^{\mathrm{(3)}}$ & $42.64^{+0.1}_{-0.1}$ & $-3.07^{+0.13}_{-0.13}$ \\ 
5.7$^{\mathrm{(4)}}$ & $42.8^{+0.16}_{-0.16}$ & $-3.11^{+0.29}_{-0.31}$  \\\hline
\end{tabular}
\begin{list}{}{}
\item[References.] (1) derived for our sample; (2)\cite{Ota2008}; (3) \cite{Ouchi2010}; (4) \cite{Ouchi2008}%; $\star$ corrected for completeness.
%(2) with IOK1 taken in account in the sample;
\end{list}
\end{table}

\textit{Sample Incompleteness.} We create a grid pattern of 15000 objects on a mock image, andrun SExtractor for different photometric apertures in double image mode, using the $g'$-band image as the measurement image. We remark that by increasing the photometric aperture size, the number of objects matching the optical criterion (Criterion \#2 in Table~\ref{tab:crit}) decreases. For an aperture of 5 pixels, we recover 75\%$\pm$0.65\%, for 10 pixels 63\%$\pm$0.6\% and for 20 pixels 46\%$\pm$0.5\%. We choose a photometric aperture of 5 pixels for the objects catalogs we used for the high redshift LAEs selection. We know then that we could miss 25\% of the objects due to the photometric aperture size we choose. This corresponds to the possibility that we missed $\sim$1.5 objects in our high redshift sample. Assuming one more object in our sample, our conclusion about the best-fit LF will not change. \\

%\begin{figure}
%\centering
%\resizebox{\hsize}{!}{\includegraphics{lf2_AC_iye_mod.pdf}}
%\caption{Best-fit Schechter function for the cumulative z$\sim$7 \La\ luminosity function. The red solid line is the best-fit z$\sim$ LF with completeness correction. Here we have included the z=6.96 confirmed LAE from \cite{Iye2006} in the determination of the best-fit Schechter parameters. Our candidates are represented as triangles (plain: not corrected for completeness, empty : with completeness correction), the z=6.96 LAE from \cite{Iye2006} is the square, \cite{Ota2010} candidates are pentagons. Also represented here is the z=6.5 \La\ LF from \cite{Ouchi2010} and the z=5.7 \La\ LF from \cite{Ouchi2008}.}
%\label{lfusiye}
%\end{figure}
%\begin{figure}
%\centering
%\resizebox{\hsize}{!}{\includegraphics{lfmodel_new.pdf}}
%\caption{Best-fit Schechter function for the cumulative z$\sim$7 \La\ luminosity function for the two cases : including or not the confirmed z=6.96 LAE in the sample.  Our candidates are represented as triangles, the z=6.96 LAE from \cite{Iye2006} is the square. Also represented here is the z=6.5 \La\ LF from \cite{Kashikawa2006}, the z$\sim$7 \La\ LF inferred by \cite{Ota2008}.}
%\label{lfmod}
%\end{figure}
\textit{Interpretation.} The previous studies presenting z$\sim$7 \La\ emitters \citep{Ota2008,Ota2010} has lead to the first spectroscopic confirmed z$\sim$7 LAE, called IOK1. Their survey covers an area of 876 square arcmin with the filter NB973 ($\Delta\lambda=200\AA$, $\lambda_{c}=9755\AA$) and reaches a 50\% completeness of  NB973=25.6 (AB, $5\sigma$) (equivalent to a flux of $1.36e^{-17}\mathrm{erg}\, \mathrm{s}^{-1}, \mathrm{cm}^{-2}$) in the SXDS (Subaru/XMM-Newton Deep Survey) and NB973=25.3 (AB, $5\sigma$) (equivalent to a flux of $1.8e^{-17}\mathrm{erg}\, \mathrm{s}^{-1}, \mathrm{cm}^{-2}$) in the SDF (Subaru Deep Field). %, corresponding to a line flux of $6 \times 10^{-18}\mathrm{erg}\, \mathrm{s}^{-1}\, \mathrm{cm}^{-2}$. 
IOK-1 has a flux of $2\times10^{-17}\mathrm{erg}\, \mathrm{s}^{-1}\, \mathrm{cm}^{-2}$. From the Table 1 of \cite{Ota2008}, we are able to obtain a lower limit for the rest-frame equivalent width of IOK1, $EW_{rest} \sim 49\AA$, using Equation ~\ref{eq:ew}. This rest-frame equivalent width is in agreement with the rest-frame equivalent limit lower limit we found for our candidate sample and presented in Table ~\ref{tab:cand}. %This object is brighter than our candidates and its equivalent width is larger than the estimates for our candidates. \\
%\textbf{but real equivalent width of this object should take in consideration that maybe flux in z band is flux continuum + some percent of flux line}\\
\cite{Ota2010} find four new photometric candidates in a survey covering the Subaru/XMM Newton Deep Survey Field with Suprime-Cam and reaching a depth limit of NB973=25.4, corresponding to 72\% completeness. These candidates are represented by pentagons in the Figure ~\ref{lfus}.\\
 %The second, shown in Figure~\ref{lfusiye}, takes in account the existence of IOK1 and includes this object in the sample for obtaining a Schechter fit. Both of these z$\sim$7 \La\ LF suggest no evolution from z=6.5 to z$\sim$7.%, and are in favor of an evolution in luminosity of $L^{*}_{z=7}\sim 0.42L^{*}_{z=6.5}$ in the case of our sample only, and $L^{*}_{z=7}\sim 0.54L^{*}_{z=6.5}$ in the case of our sample combined with \cite{Iye2006} point.\\
% in luminosity from $\mathrm{log}(L^{*})$=42.20 to $\mathrm{log}(L^{*})$=42.26.
%In our second case, i.e. taking in account the z=6.96 spectroscopic LAE, we remarks a slight evolution, especially in density from  that the faint end is strongly affected by the completeness correction $\mathrm{log}(\Phi^{*})$=2.80 to $\mathrm{log}(\Phi^{*})$=2.02.
%In Figure~\ref{lfmod}, we compare our photometric sample to \cite{Ota2008} assumption that the luminosity function evolves from z=5.7 to  z$\sim$7 by a factor of $L^{*}_{z=7}=0.58L^{*}_{z=5.7}$, with $L^{*}_{z=5.7}=1.08.10^{43}\mathrm{erg}\, \mathrm{s}^{-1}$ \citep{Shimasaku2006}. 
%Both of this possible z$\sim$7 \La\ LF do not fit our data.\\

%\cite{Ota2008}, by assuming that the LAEs and LBGs show a similar evolution history, can infer that the \La\ LF derived from LAEs evolves as the rest-frame UVLFs obtained from LBGs. 
From \cite{Yoshida2006}, \cite{Ota2008} estimated therefore a possible z$\sim$7 \La\ LF with a pure luminosity evolution of  $L^{*}_{z=7}=0.58L^{*}_{z=5.7}$, with $L^{*}_{z=5.7}=1.08.10^{43}\mathrm{erg}\, \mathrm{s}^{-1}$ \citep{Shimasaku2006}. This inferred z$\sim$7 \La\ LF predicts fewer LAEs than seen in our photometric candidate sample. Confirmation of 1-2 of our candidates would approximately match the prediction in \cite{Ota2008} and would modestly exceed their measured number density.\\
%does not agree with our photometric candidates sample. It confirms their idea that, if LAEs are strongly related to LBGs, the neutral hydrogen fraction in the IGM is possibly higher at z$\sim$7 than at z$\sim$5.7, and this difference can cause the attenuation of \La\ lines of high redshift LAEs.\\

We show in Figure~\ref{lfus} the cumulative z$\sim$7 LAEs LF obtained after correcting our points for the aperture and the detection completeness. This completeness correction has been applied by number weighting according to the NB9860 magnitude. The best-fit parameters do not vary significantly before and after correcting from the completeness, as seen in Figure~\ref{lfus} between the black solid line and the red solid line (without and with the completeness correction, respectively). \\
By considering only our photometric candidate sample, we do not observe any strong $L^{*}$ or $\Phi^{*}$ evolution between z=5.7 and z$\sim$7, and therefore contradict a possible $L^{*}$ evolution between z=6.5 and z$\sim$7.
If none of our candidates is a real z$\sim$7 LAE, we can then put an upper limit on the z$\sim$7 \La\ LF, which will help constrain the neutral fraction of the IGM.
%more evolution in density than in luminosity.

\section{Conclusions}
We observed 465 $\mathrm{arcmin}^{2}$ from the COSMOS field using the narrow-band imaging technique on Magellan/IMACS with the $NB9680$ filter, in order to target the \La\ line at z$\sim$6.96. We obtained a comoving volume of $\sim 72000 Mpc^{3}$.
%We observed during six nights, between 2009 and 2010, 465 $\mathrm{arcmin}^{2}$ in the COSMOS field with the NB9680 filter, corresponding to targeting the \La\ line at z$\sim$6.96.
 After applying our selection criteria and verifying that our selection was not contaminated by low-redshift emitters and transient objects, we obtain a sample of six z$\sim$6.96 LAEs. From this photometric sample, we are able to infer a possible z$\sim$6.96 \La\ Luminosity Function. We find no evolution in luminosity function from z=6.5 to z$\sim$6.96, if a majority of our sources are confirmed.\\
It is now crucial to obtain spectroscopic follow-up observations to reveal the real nature of these objects and establish a firm conclusion on the z$\sim$6.96 \La\ Luminosity Function.
\vspace{0.5cm}

\acknowledgments
The authors would like to thank the referee for very useful comments, Las Campanas Observatory staff, the IMACS team and the National Science Foundation.
%-----------------------------------------------------\\
\bibliographystyle{apj}
\bibliography{hibo1230_revised6}
\end{document}